**Abstract** Microsphere-assisted imaging emerged as a surprisingly simple way of achieving optical super-resolution. In this work, by using movable polydimethylsiloxane (PDMS) thin films with embedded high-index ($n\sim 2$) barium titanate glass microspheres a sample scanning capability was developed, thus removing the main limitation of this technology based on its small field-of-view. It is shown that such thin films naturally adhere to the surface of nanoplasmonic structures so that the tips of embedded spheres experience the object near-fields. Based on rigorous criteria, the resolution $\sim\lambda/6$-$\lambda/7$ (where $\lambda$ is the illumination wavelength) is demonstrated for such thin films. After surface lubrication with isopropyl alcohol (IPA), the thin films become non-sticky and can be translated along the surface to provide precise alignment of microspheres with the objects of studies. It is shown that the resolution gradually increases beyond the diffraction limit as the lubricant evaporates. Such thin films can become a standard component in super-resolution imaging of nanostructures and biological objects.

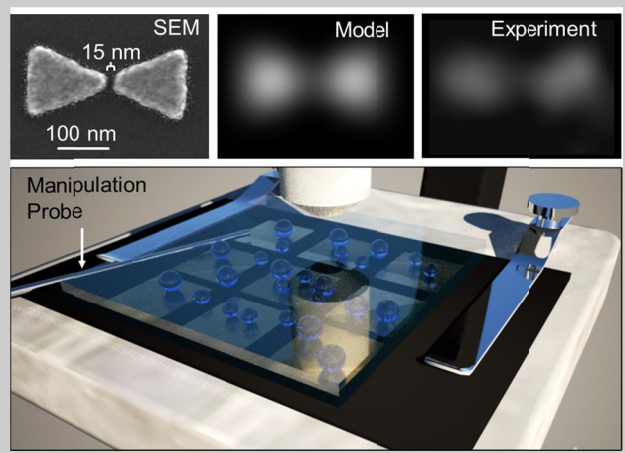

# Movable thin films with embedded high-index microspheres for super-resolution microscopy

**Kenneth W. Allen**[1,2,*], **Navid Farahi**[1,2], **Yangcheng Li**[1], **Nicholaos I. Limberopoulos**[2], **Dennis E. Walker Jr.**[2], **Augustine M. Urbas**[3], **Vladimir Liberman**[4], and **Vasily N. Astratov**[1,2*]

## 1. Introduction

Imaging by dielectric microspheres emerged as a surprisingly simple way of obtaining super-resolved images of nanoplasmonic structures [1-9] and biological objects [5, 7]. Initially, the method has been demonstrated for micron-scale low-index ($n_s$=1.46) silica spheres in air [1], later it was extended to silica spheres semi-immersed in a liquid [2], and, finally, it has been developed [3,4] for high-index ($n_s$>1.8) spheres totally submersed in a liquid. A microsphere (or cylinder) plays the part of a "magnifying lens" placed in contact with the object. The microsphere creates a magnified virtual image that can be viewed by a standard microscope at a certain depth inside the structure.

A surprising aspect of the virtual images produced is the incredible level detail of the tiny features of the objects. Many groups reported extraordinarily high resolution of various nanoscale objects spanning the range from $\lambda/6$ to $\lambda/17$ [1-9]. It should be noted that different resolution criteria have been used that complicates the direct comparison of these results. The present work contains an in-depth analysis of these resolution criteria. However, irrespective of the chosen method, all these studies have demonstrated that the observed resolution far exceeds the diffraction limit. According to classical definition, two point objects with the same brightness are resolved if they are separated by a certain distance, $d \geq K\lambda/\mathrm{NA}$, where $K$=0.5, 0.61, 0.473, and 0.515, according to Abbe [10], Rayleigh [11], Sparrow [12], and Houston [13] criteria, respectively. Here, the numerical aperture of the imaging system is represented by NA=$n_o \times \sin\theta$, where $n_o$ is the object-space refractive index and $\theta$ is the half of the objective's acceptance angle.

---

[1] Department of Physics and Optical Science, Center for Optoelectronics and Optical Communications, University of North Carolina at Charlotte, Charlotte, NC 28223-0001, USA
[2] Air Force Research Laboratory, Sensor Directorate, Wright-Patterson AFB, OH 45433 USA
[3] Air Force Research Laboratory, Materials and Manufacturing Directorate, Wright Patterson AFB, OH 45433 USA
[4] Lincoln Laboratory, Massachusetts Institute of Technology, Lexington, Massachusetts 02420, USA
* Corresponding authors: e-mails: kenneth.w.allen.jr@gmail.com; astratov@uncc.edu

It is natural to suggest that the super-resolution capability of microspheres stems from the evanescent field enhancement provided by microspheres or cylinders [14, 15]. It can also be related to the ultra-sharp focusing properties of microspheres termed "photonic nanojets" [16]. However, generally, the microsphere can produce a near-field focal spot with lateral resolution slightly beyond $\lambda/2n_s$, which is also the lateral resolution limit of the dielectric microsphere and solid immersion lens [17]. If the super-resolution imaging capability of microsphere is directly determined by its focusing properties, when the resolution is supposed to be limited at the level $\sim\lambda/4$ for $n_s$=2 which is significantly lower resolution than the experimental values varying from $\lambda/6$ to $\lambda/17$ [1-9].

It seems plausible to suggest that the super-resolution capability of microspheres can be explained by collecting the high spatial frequencies of an object encoded in its optical near-fields which are typically lost when conventional lenses are used. The detailed mechanisms are, however, debated in the literature including a potentially important role of surface plasmon polaritons [18] and surface states [19]. It has also been shown that the internal resonances in microspheres can facilitate sharper focusing [20, 21]. Two-point resolution criterion has been tested with microspheres using classical imaging theory [22, 23]. The super-resolution effect has not been revealed [22, 23] indicating that other mechanisms such as surface roughness enhanced plasmon resonance could be involved.

Microsphere-assisted imaging can have a tremendous impact on microscopy, however a drawback of this technology is its narrow field-of-view (FOV) limited at $\sim D/4$, where $D$ is the sphere diameter [4]. It has been experimentally demonstrated that this method provides super-resolution for smaller spheres with the size parameter $q=\pi D/\lambda<100$ [1, 4]. This means that in practice FOV is limited by few microns. Larger areas can be inspected by locomotion of microspheres using the optical tweezers [24, 25] or micromanipulation [6].

In this work, we developed a different approach to this problem based on the incorporation of BTG spheres in intrinsically flexible, mechanically robust and optically transparent polydimethylsiloxane (PDMS) thin films (or "coverslips"), as illustrated in Fig. 1. The key element of the design of coverslips is a planar array of high-index BTG microspheres with a broad range of diameters 2<$D$<53 $\mu$m held in nanometer-scale proximity to the bottom surface of the coverslips. Once the coverslip is attached to a nanoplasmonic structure, the tips of microspheres can experience the object near-fields leading to the possibility of super-resolution imaging.

We show that the PDMS coverslips are naturally adherent to various substrates providing imaging through the embedded microspheres with $\sim\lambda/6$-$\lambda/7$ resolution. Our initial results are presented in [26, 27]. A resolution limited at $\sim\lambda/4$ by non-movable coverslips was reported in [28]. We use a rigorous definition of the resolution limit for finite-size objects based on the convolution of the point-spread-function (PSF) with the object's intensity distribution. After surface lubrication with isopropyl alcohol (IPA), the coverslips temporarily lose their adherent properties, and they can be translated along the surface. The ability to simultaneously capture images through the two-dimensional (2-D) array of spheres during widefield microscopy allows precise alignment of microspheres with the objects of studies. We show that just after lubrication the resolution is determined by the diffraction limit. However, as the lubricant evaporates, the resolution gradually increases beyond the diffraction limit.

## 2. Fabrication: Nanoplasmonic Objects and Coverslips with Embedded Spheres

### 2.1. Nanoplasmonic Objects

As objects for imaging, we used arrays of golden dimers and bowties illustrated by scanning electron microscopy (SEM) images in Figs. 1(a) and 1(b), respectively. The dimers in Fig. 1(a) were fabricated at Air Force Research

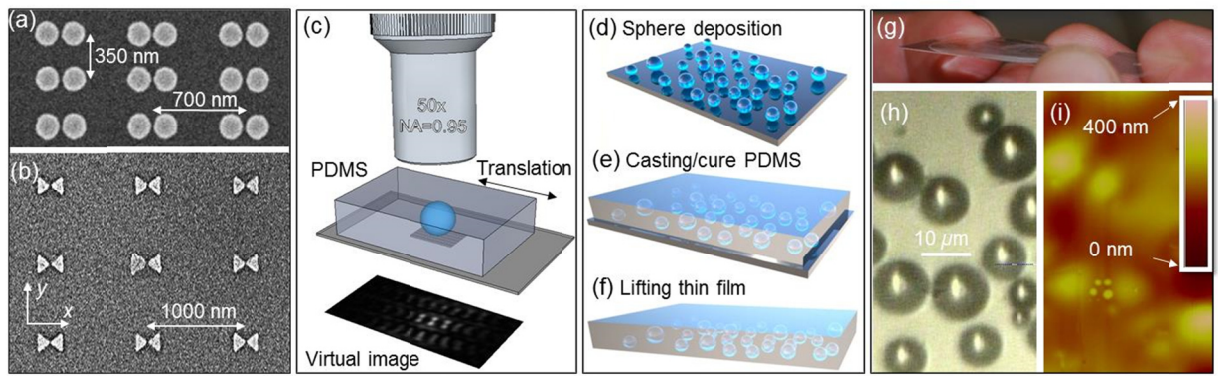

**Figure 1** (a,b) SEM of the 2-D arrays of golden dimers and bowties, respectively. (c) Schematic of the setup. (d-f) Step-by-step fabrication of coverslips with embedded spheres. (g) Photograph of the coverslips. (h) Microscopic image of the coverslip's bottom surface. (i) Bump mapping at the bottom surface by atomic force microscopy.

Laboratory (AFRL) [8]. They were assembled in 2-D arrays with periods of 700 nm and 350 nm in *x* and *y* directions, as illustrated in Fig. 1(a). The dimers in Fig. 1(a) consisted of two 185 nm gold nanocylinders with 10 nm Ti/40 nm Au heights and 65 nm edge-to-edge separations. These parameters varied in different arrays. The SEM image in Fig. 2(a) represents dimer formed by 120 nm nanocylinders with 60 nm separation (along *x*-axis).The fabrication was performed on a sapphire substrate by an electron beam lithography, metal evaporation, and liftoff process [8].

The bowties in Figs. 1(b) and 2(e) were fabricated at Massachusetts Institute of Technology, [29]. They were assembled in 2-D square arrays with 1 *μ*m period. The bowties were produced by an advanced nanostencil lithography process that provides sub-15 nm resolution even for 40-nm thick structures. The patterns in the ultrathin freestanding silicon nitride membrane were converted through the apertures in the stencil mask into metallic nanostructures on the target substrate with 4 nm Ti/35 nm Au heights. The proximity between the stencil and substrate was controlled by a sacrificial poly-methyl methacrylate layer. The fabrication was performed on a silicon substrate.

### 2.2. Barium Tinanate Glass Microspheres

The BTG spheres used in this work have been produced by Mo-Sci Corp. in two modifications [30]. The spheres with larger content of barium have index $n_s$~1.9 and the spheres with larger content of titanium have index $n_s$~2.1 for the red portion of the visible region. In most of our experiments we used the former modification due to better overall quality of microspheres. Imaging was performed by a scanning laser confocal microscope Olympus LEXT-OLS4000 operating at $\lambda$=405 nm where the BTG microspheres with $D\leq50$ *μ*m have negligible absorption. Due to glass dispersion, the index of BTG spheres with the excess barium content was estimated to be $n_s$~2.0 at $\lambda$ = 405 nm providing index contrast with PDMS ~1.4 that is rather close to that for silica spheres in air. Similar to liquid-immersed BTG spheres [8], the BTG microspheres embedded in PDMS are expected to have an inherent resolution advantage over silica spheres in air due to their higher object-space index.

### 2.3. PDMS Coverslips with Embedded Spheres

To fabricate the coverslips with embedded spheres we used a three-step process [31] schematically illustrated in Figs. 1(d-f). This process bares some similarity with previously developed technology of embedding polystyrene microspheres in the PDMS membranes for applications in projection lithography [32]. First, the BTG spheres were deposited on the surface of a microscope slide where they formed a disordered monolayer. Then, a PDMS (*Sylgard®* 184 Silicone Elastomer produced by Dow Corning Corp., 10 to 1 mix ratio) layer was cast over the spheres. After filling the cavities, formation and floating the microbubbles, a continuous layer with 200-300 *μ*m thickness containing microspheres was produced. It was cured at 90°C for one hour. After this treatment, it was transformed into a tough, elastomeric thin layer. The photograph of this layer at the top of the microscope slide is presented in Fig. 1(g). Finally, this thin layer was lifted from the substrate with tweezers and used as a coverslip in super-resolution studies. The adhesion of the PDMS to the wafer complicates the separation of coverslips from the substrate. In principle, the adhesion to the substrate can be reduced by the silanization [32]. We did not use any surface chemical treatments and were still able to safely separate

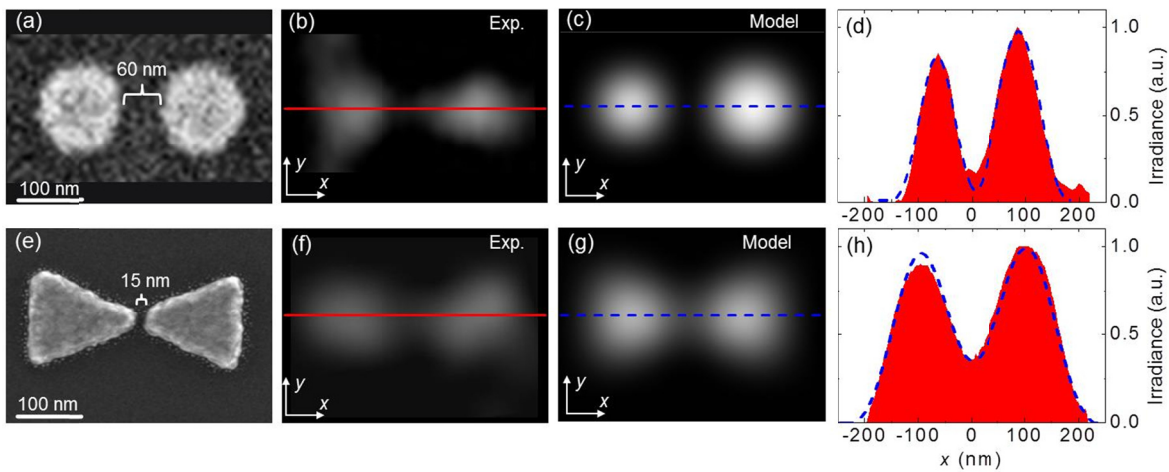

**Figure 2** (a,e) SEM image of the Au dimer and bowtie, respectively. The dimer is formed by 120 nm nanocylinders with 60 nm separation. (b,f) Images of dimer and bowtie in (a,e) obtained through the 5 μm and 53 μm BTG microsphere, respectively. (c,g) Calculated convolutions of the idealized (perfect) object of dimer and bowtie (with the same dimensions as in (a,e)) with the 2-D PSFs with FWHMs of $\lambda$/7 and $\lambda$/5.5, respectively. (d,h) Comparison of the measured (red background) and calculated (dashed blue curves) irradiance profiles through the cross-section of the Au dimers (*x*-axis).

coverslips with areas of 1×1 cm$^2$ from the microscope slides. These coverslips can be handled by the tweezers and applied to the surface of investigated sample.

The microspheres remained embedded in the coverslip after it has been separated from the surface of the wafer. Inside such coverslips, most of the spheres were separated from the lower interface by a nanometer-scale distances (<100 nm). A small fraction of microspheres were protruding through the PDMS. The bump mapping at the bottom surface by atomic force microscopy (AFM), see Fig. 1(i), shows a strong correlation of the bump positions with the locations of embedded spheres illustrated by a microscope image in Fig. 1(h). The lower interface of the coverslip was found to have a relatively smooth surface with the bumps' heights limited at ~200 nm.

## 2.4. Use of Coverslips without Lubrication

For super-resolution imaging applications of coverslips with embedded microspheres, we need a combination of two somewhat contradicting material properties. The first property is to have a sufficiently strong adherence to the investigated surface which would allow for many embedded microspheres to come sufficiently close to the surface. The second property is an ability to eventually detach the coverslips from the sample with a minimal damage to the both contacting surfaces.

Due to the weak bond (physical adsorption) between the PDMS coverslips and underlying surface of glass or semiconductor substrates [33], such films can be reversibly attached to the investigated surfaces. Immediately after such attachment, an interference pattern can be observed indicating the several micron gaps between the coverslip and surface. Without any additional pressure, the interference fringes gradually vanish on a time scale of several seconds, indicating that the gaps reduced to subwavelength dimensions. We were not able to monitor further reduction of the gaps, but based on our observation of the initial dynamics of this process, we believe that several minutes should be sufficiently long time to assume that the lower interface of the coverslip is located in a nanoscale vicinity of the investigated surface. It should be noted that in principle, the adherence to the substrate can be reduced by silanization [32] or increased by oxidization followed by the use of pressure [34]. We found that the PDMS coverslips fabricated without any additional surface treatment are easily separable from the investigated samples. They can be completely peeled off from the sample resulting in free-standing PDMS films which can be multiply used for super-resolution imaging on different samples.

## 3. Definition of the Optical Super-Resolution

It has been well established that the microsphere-assisted imaging can make visible fine features of the objects which cannot be resolved by the best diffraction-limited microscope objectives with NA approaching unity [1-9]. Defining, however, the numerical value of the optical super-resolution has been a somewhat controversial issue in these studies. The fundamental reason for this difficulty is based on the fact that the textbook resolution criteria are formulated for idealized point objects [35, 36]. For diffraction-limited systems operating in visible or near-visible regimes many structures would qualify as almost perfect point objects including nanoscale apertures, nanoplasmonic clusters, individual quantum dots or dye molecules [37, 38]. For systems possessing optical super-resolution, however, the realization of point objects is more challenging task. This is especially true for imaging systems operating in reflection or transmission modes since the objects with nanoscale dimensions produce low intensity images.

To circumvent this problem, researchers often select a different path based on using larger-scale arrays containing features with recognizable shape such as periodic Blu-ray® disks [1 – 9], nano-pores in fishnet metallo-dielectric films [1, 5], star shapes [1], nanoplasmonic dimers [4] or more complicated clusters [9]. In such cases, the resolution claims are often based on observability of "minimal discernible feature sizes". Different features such as widths of the stripes, diameters of nanopores, edge-to-edge separations in dimers and clusters have been studied by microsphere-assisted technique that resulted in a broad range of resolution claims from $\lambda$/6 to $\lambda$/17 [1, 4-7, 9].

Although these results have been extremely important for the advancement of the area of super-resolution imaging by microspheres, we would like to stress that an extreme caution should be exercised with using the criteria of observability of "minimal discernible feature sizes". Below, we illustrate that it can result in greatly overestimated resolution compared to that obtained by a standard procedure of convolution with PSF.

As an object, we consider arrays of dimers and bowties illustrated by SEM images in Figs. 1(a) and 1(b), respectively. Individual dimers and bowties used for imaging are illustrated in Figs. 2(a) and 2(e), respectively. Conventional or confocal microscopy without microspheres performed with the 100×(NA=0.95) objective lens at $\lambda$=405 nm does not allow resolving the internal structure of individual dimers or bowties illustrated by SEM images in Figs. 2(a) and 2(e), respectively. In contrast, the microsphere-assisted imaging demonstrates significantly improved resolution. The virtual image of a dimer in Fig. 2(a) obtained through the 5 μm BTG sphere embedded in the PDMS coverslip is shown in Fig. 2(b). The virtual image of a bowtie in Fig. 2(e) obtained through the embedded 53 μm BTG sphere is shown in Fig. 2(f).

Irradiance profiles along the *x*-axis of dimers and bowties are illustrated using red as background color in

Figs. 2(d) and 2(h), respectively. The saddle-to-peak ratios of these profiles, 0.16 and 0.35, respectively, are significantly smaller than that assumed in various classical definitions of resolution of two point sources [36, 37].

An attempt to define the resolution based on observation of minimal discernable features can lead to misleading results if, for example, we interpret the saddle point in Fig. 2(h) as a manifestation of resolution of ~15 nm gap in bowties which would imply the resolution in excess of ~$\lambda/27$. In fact, as we show below, the image reconstruction with the Gaussian PSF with the width ~$\lambda/5.5$ allows obtaining a high-quality fit to the experimental results presented in Figs. 2(g) and 2(h). This means that the resolution in this case is ~$\lambda/5.5$, and it can also be shown that it practically does not depend on the existence of the ~15 nm gap.

We treated the super-resolved images based on analogy with the classical theory [35] where the image, $I(x, y)$, is considered as a convolution of a diffraction-limited PSF and the object's intensity distribution function, $O(u, v)$. This can be expressed in the standard integral form:

$$I(x,y) = \iint_{-\infty}^{\infty} O(u,v) PSF(u - x/M, v - y/M) du dv, \quad (1)$$

in which the integration is performed in the object plane where the coordinates $(x_o, y_o)$ are linearly related to the image plane via the magnification $M$ as: $(x_o, y_o)=(x_i/M, y_i/M)$. We used a Gaussian function for PSF$(x_o, y_o)$ with the full width at half maximum (FWHM) being a fitting parameter. Based on the Houston criterion, fitted values of FWHM in the object plane were considered a resolution of the system.

In a standard textbook approach [35, 36] the PSF width is diffraction-limited. To permit the description of super-resolution we allowed subdiffraction-limited FWHM values, however we kept the same basic Eq. (1) for image reconstruction. Thus, this is a phenomenological approach which does not specify the mechanism of super-resolution. However, it allows determining the super-resolution value in a way which is consistent with the resolution definitions widely accepted in diffraction-limited optics. Similar approach has been used previously to provide 1-D treatment with rectangular functions [8]. In the present work, we generalized this approach for 2-D PSF and for objects with arbitrary shape (such as bowties) that allowed us to directly calculate the images and to compare them with experimentally observed 2-D images. Such images calculated with 2-D PSFs with FWHMs ~$\lambda/7$ and ~$\lambda/5.5$ are presented in Figs. 2(c) and 2(g) for dimers and bowties, respectively. Very good agreement with experimental images in Figs. 2(b) and 2(f) was found. The calculated intensity profiles along $x$-axis are represented by blue dashed lines in Figs. 2(d) and 2(h). They also demonstrate very good agreement with the experimental intensity profiles illustrated using red as background color in Figs. 2(d) and 2(h), respectively. These results show that the super-resolution available with elastomeric PDMS coverslips without lubrication is ~$\lambda/6$-$\lambda/7$ which is similar to the resolution observed [4, 8] for liquid-immersed BTG spheres.

## 4. Locomotion of Lubricated Coverslips

To align embedded spheres with various objects we developed a technique based on lubrication and locomotion of the entire PDMS thin film containing massive number of BTG spheres, as illustrated in Fig. 3. The lubrication was provided by using easily evaporable liquid such as isopropyl alcohol (IPA) with index 1.37, as schematically shown in Fig. 3(d). The locomotion of the coverslip was performed by a tapered stainless steel microprobe inserted in the PDMS, as also illustrated in Fig. 3(d). The probe was connected with a hydraulic micromanipulation controller providing ~1 $\mu$m precision of translation. Taking into account that each sphere has FOV~$D/4$ [4], such precision was sufficient for aligning various spheres with the object of interest. The locomotion of the coverslip was controlled by visual inspection using the FS70 Mitutoyo microscope with peak illumination at $\lambda$~550 nm and 20×(NA=0.4) objective lens.

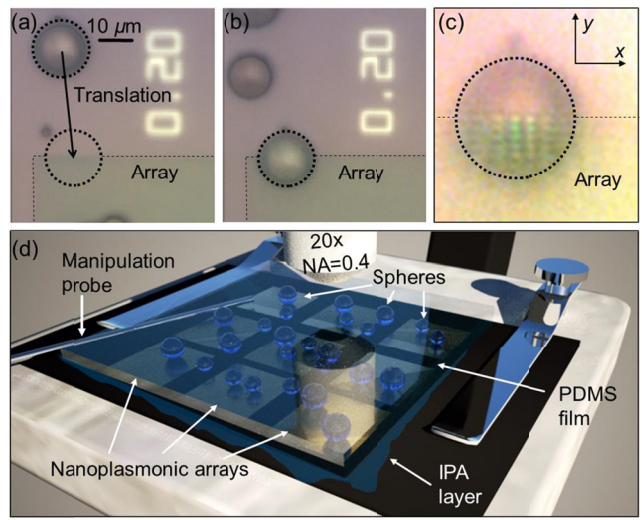

**Figure 3** Translation of the coverslip lubricated with IPA. (a) The embedded BTG sphere is ~40 $\mu$m away from the edge of an Au nanoplasmonic array. (b) The same sphere is at the border of array. (c) By changing the depth of focus, the dimers are seen near the array's edge. (d) The motion of the coverslip is provided by a metallic probe inserted in PDMS and connected to a micromanipulator.

The translation and imaging through the coverslip immediately after its lubrication and application to the sample surface are illustrated in Figs. 3(a-c). Initially, the 16 μm BTG microsphere is located approximately 40 μm away from the border of the nanoplasmonic array, as shown in Fig. 3(a). As a result of the shift of the coverslip by ~40 μm, this sphere is aligned with the edge of the array, as shown in Fig. 3(b). The virtual image of the array was obtained by the 20×(NA=0.4) objective lens by focusing deeper in the structure, as shown in Fig. 3(c). The diffraction-limited resolution of the microscope objective can be estimated as $d=0.515\lambda/NA\sim700$ nm. However, experimentally, both the 700 nm and 350 nm periods of the arrays illustrated in Fig. 1(a) can be discerned in the virtual image presented in Fig. 3(c). It can be explained by the fact that the microsphere, not the microscope objective, defines the angle $\theta$ of the microsphere-assisted imaging [8] and higher effective NA values can be realized. The internal structure of the dimers, however, cannot be resolved through the microspheres even by using the 100×(NA=0.9) microscope objective lens.

Achieving the super-resolution in such situation requires reducing the thickness of the IPA layer to bring the embedded spheres closer of the object. This requires investigation of the gaps ($g$) separating the coverslip from the surface in the course of IPA evaporation, see the geometry in Fig. 4(a). We used for this purpose the lateral image magnification ($M$) which in the limit of geometrical optics is related to the gap by the following equation:

$$M(n',r,g) = \frac{n'}{2(n'-1)\left(\frac{g}{r}+1\right)-n'}, \quad (2)$$

where $r$ is the radius of the sphere, and $n' = n_{sp}/n_o \sim 2.0/1.37 \approx 1.46$ is the refractive index contrast between the sphere and object space. The dependence of $M$ on the normalized gap ($g/r$) calculated using Eq. (2) is illustrated for $n'=1.5$ by the black line in Fig. 4(b). In the limit of $g \ll r$ it predicts $|M| \sim |n'/(2-n')| \sim 3$ in agreement with the previous results [1, 4] obtained for the virtual image of an object located at the sphere surface.

The dynamics of magnification in the course of the IPA evaporation was studied by the scanning laser confocal microscope Olympus LEXT-OLS4000 with the 100×(NA=0.95) objective lens at $\lambda = 405$ nm, as illustrated in Figs. 4(d-f). For these studies, we selected a rectangular array of Au dimers formed by the 185 nm cylinders with 65 nm edge-to-edge separations illustrated by SEM image in Fig. 4(c).

The magnification ($M$) of the virtual image created by the 16 μm sphere embedded in a PDMS coverslip was determined in comparison with the real image of the surface of the structure obtained outside the microsphere [4]. The experimental results demonstrate a reduction of magnification from $M \approx 5.4$ measured in the first minute after application of the coverslip (Fig. 4(d)) to $M \approx 3.1$ after 72 hours (Fig. 4(f)). It should be noted that the $M$ measurements are more precise in confocal mode compared to conventional microscopy because of the better in-depth resolution. This is determined by much more compact axial FWHM of PSF in confocal microscopy available for the minimum detector apertures [39]. Still, the small deviations from the optimal focusing depth lead to marked $M$ variations which causes the measurement error about 5-6% illustrated by the vertical error bars in Fig. 4(b). The measured $M$ values were transported on the calculated $M(g/r)$ dependence to estimate the gaps. Originally, for the first image obtained, Fig. 4(d), the gap is ~1.8 μm. After 10 minutes of IPA evaporation, Fig. 4(e), the gap was reduced to ~1 μm. Then after 72 hours, it appears that the IPA layer has almost completely evaporated and the gap has approached close to zero (~100 nm), Fig. 4(f), according to our fit, as shown in Fig. 4(b). Determination of more precise steady-state value of nanoscale gaps is made difficult by the errors of magnification measurements.

The resolution studies were performed in parallel with the magnification measurements using the same array of dimers and same 16 μm BTG sphere as in Figs. 4(d-f). These dimers formed by 185 nm cylinders with 65 nm edge-to-edge separations were barely resolved within first minute after lubrication and application of the coverslip (Fig. 4(d)). However, the structure of individual dimers became much better visible in the images obtained later in the course of evaporation of IPA layer, even despite the fact that the magnification was gradually decreasing (Figs. 4(e,f)). This is better seen in the magnified images of this dimer obtained after drying for 10 min and 72 hours, as illustrated in Figs. 5(a) and 5(c), respectively. Irradiance profiles along the $x$-axis of dimers are illustrated using red as background color in Figs. 5(b) and 5(d), respectively. The saddle-to-peak ratio decreased from 0.34 to 0.23 and the intensity peaks became narrower in the profile presented in Fig. 5(d) compared to that in Fig. 5(b), indicating improved system resolution with time.

Using a PSF-based image fitting procedure described in Section 3, we calculated intensity profiles along $x$-axis represented by blue dashed lines in Figs. 5(b) and 5(d). They demonstrate very good agreement with the

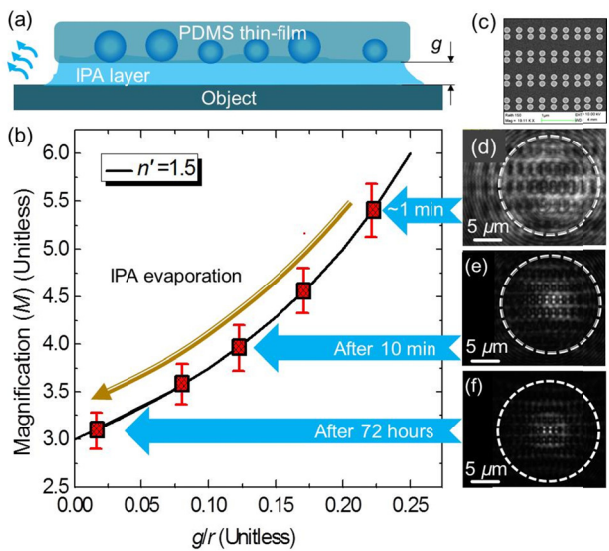

**Figure 4** Decrease in magnification ($M$) as a consequence of reducing the gap ($g$) between the embedded 16 μm BTG microsphere and a dimer array. (a) Schematic illustration of the IPA layer evaporation. (b) Dynamical measurements of $M$ (points) transported on the $M(g/r)$ dependence (black line) for $n'=1.5$. (c) SEM image of the dimer array. Virtual images through the same sphere (d) after lubrication, (e) after 10 min, (f) after 72 hours.

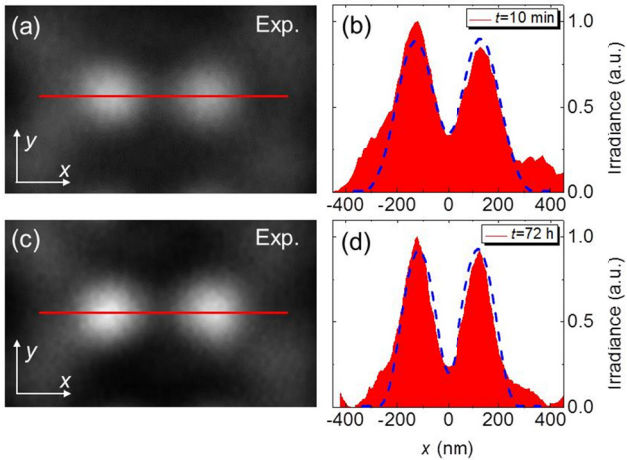

**Figure 5** (a) Image of an Au dimer formed by 185 nm cylinders with 65 nm edge-to-edge separations obtained through the 16 µm embedded BTG microsphere after 10 minutes of the IPA evaporation. (b) Comparison of the measured (red background) and calculated (dashed blue curves) irradiance profiles through the cross-section of the Au dimers (*x*-axis). (c,d) Same as (a,b), respectively, but after drying for 72 hours.

experimental intensity profiles at the FWHW of the Gaussian PSF fitting function at $\sim\lambda/4$ and $\sim\lambda/5.5$ in Figs. 5(b) and 5(d), respectively. These results show that the super-resolution provided by the embedded BTG microspheres gradually increases as the IPA layer evaporates. The resolution values obtained after 72 hours ($\sim\lambda/5.5$) almost reached the level of resolution obtained in structures without lubrication ($\sim\lambda/6$-$\lambda/7$).

## 5. Conclusions

In conclusion, due to simplicity and broad range of wavelengths, super-resolution imaging by dielectric microspheres has the potential to provide a strong impact on microscopy. There have been many super-resolution claims in this area spanning the range from $\sim\lambda/6$ to $\sim\lambda/17$ [1-9]. Although the technique provides images of nanoscale objects with a superior quality compared to the best diffraction-limited systems, the quantitative definition of the super-resolution can be a controversial issue.

In this work, we showed that the observation of minimal discernible features in the optical images of finite objects and associating the size of these features with the optical resolution of the system is not directly related to the textbook definitions of optical resolution. For finite objects, this procedure can lead to overestimated resolution. We showed that more consistent way of defining optical super-resolution is based on a standard procedure of convolution with the point spread function widely used in diffraction-limited optics. By generalizing this approach for super-resolved images produced by microspheres we established that the resolution $\sim\lambda/6$-$\lambda/7$ can be systematically observed on metallic arrays of nanoscale dimers and bowties.

Currently, the broad applications of this technology are impeded by relatively limited field-of-view ($\sim D/4$) offered by individual spheres. The established ways of overcoming this problem are optical tweezers [25] and locomotion [6] of individual spheres.

In this work, we embedded hundreds of high-index BTG microspheres inside transparent elastomeric PDMS slabs. Such slabs or coverslips can be considered as a novel optical component for super-resolution microscopy. It can be used in two different ways: i) without lubrication, when in just few seconds the coverslip gets adhered to various surfaces providing up $\sim\lambda/6$-$\lambda/7$ resolution, and ii) with liquid lubrication, when the coverslips becomes non-sticky and it can be translated along the surface to align different spheres with various objects. We showed that in the second method the resolution initially drops to diffraction-limited level. However, as the liquid lubricant evaporates, the gap between the objects and the coverslip surface (with the near-surface spheres) is reduced to a nanometric scale providing super-resolution imaging through the microspheres.

**Acknowledgements.** The authors gratefully acknowledge support from U.S. Army Research Office through Dr. J. T. Prater under Contract No. W911NF-09-1-0450 and DURIP W911NF-11-1-0406 and W911NF-12-1-0538. This work was also supported by Center for Metamaterials, an NSF I/U CRC, Award No. 1068050. Also, this work was sponsored by the Air Force Research Laboratory (AFRL/RYD, AFRL/RXD) through the AMMTIAC contract with Alion Science and Technology and the MCF II contract with UES, Inc. The Lincoln Laboratory portion of this work was sponsored by the Department of the Air Force under Air Force Contract #FA8721-05-C-0002. Opinions, interpretations, recommendations and conclusions are those of the authors and are not necessarily endorsed by the United States Government.